# A Game Theoretic Setting of Capitation Versus Fee-For-Service Payment Systems

Allison Koenecke
*Institute for Computational & Mathematical Engineering*
*Stanford University*
Stanford, CA, USA
koenecke@stanford.edu

*Abstract*—We aim to determine whether a game-theoretic model between an insurer and a healthcare practice yields a predictive equilibrium that incentivizes either player to deviate from a fee-for-service to capitation payment system. Using United States data from various primary care surveys, we find that non-extreme equilibria (i.e., shares of patients, or shares of patient visits, seen under a fee-for-service payment system) can be derived from a Stackelberg game if insurers award a non-linear bonus to practices based on performance. Overall, both insurers and practices can be incentivized to embrace capitation payments somewhat, but potentially at the expense of practice performance.

*Keywords—healthcare costs, game theory, proactive healthcare, health care capitation, fee-for-service*

## I. Introduction

Capitation and fee-for-service (FFS) payments are two contrasting systems to pay healthcare practices. Under the capitation payment system, a fixed payment is made to the practice for each enrolled patient, per time period (the practice absorbs cost or surplus); under FFS payments, the practice is paid for each of the specific services delivered to a patient (the insurer absorbs cost or surplus). Capitation payments are often theorized to be useful in shifting primary care toward proactive team and nonvisit care, which in turn may lead to lower hospitalization rates for patients due to the shift towards preventative care. However, there is minimal literature demonstrating such objective findings, and little history of capitation payment enactment in the US [1].

The differences between potential care stemming from these two payment systems have led to vigorous debate within the US healthcare system; in particular, it is unclear whether FFS payments should be moved to capitation payments. One reason for the reluctant shift toward capitation payments is the absence of proper incentive structures that adequately reward both the insurers and practices involved. Prior work, based on payment simulations, has shown that high levels of capitation payments would be necessary for a resulting change in primary care [2]. The current economics literature has modeled insurer-practice networks through a competition and demand estimation lens, and shown that providers bear the most burden of a cost increase [3]. It has also shown that consumer welfare is negatively impacted by restricting choice of practice [4]. However, the current literature lacks studies on the relationship between insurers and practices with regard to capitation and FFS payments. Rather, prior work focuses on the effects of capitation and FFS payment systems on patients rather than a broader segment of the healthcare system.

In contrast, we use a novel game-theoretic approach in setting the share of patients seen under capitation payments, which allows us to directly measure the potential shift from FFS to capitation payments as a result of insurer-practice competition. Our work aims to fill the gap in the literature regarding the split between FFS and capitation payment system choice in the US, from both the insurer and practice perspectives. We are motivated by the existence of patients who may gain higher utility from using either FFS or capitation payments; as such, it is worthwhile to explore whether the current system allows for the operation of both payment systems regardless of a specific patient's preference. We are able to determine that there is a functional system in which both FFS and capitation payment modes co-exist, while still incorporating quality of patient care in our holistic model.

Specifically, we model the insurer as a party with the ability to set the fraction of patients $f_1$ under an FFS payment system (as opposed to a capitation payment system). In response, we model the practice as a party that can set the fraction of patient visits $f_2$ conducted under an FFS model (i.e., visits without proactive team or nonvisit care). This can be played as a Stackelberg game [5] wherein the first player, the insurer, sets the value of $f_1$ so that the insurer's net cost is minimized. In response, the practice sets $f_2$ so that the practice's total revenue is maximized. Further, we introduce a performance-based bonus mechanism [6] for insurers to either reward or penalize practices.

In practice, fraction-setting of $f_1$ and $f_2$ can be done in several ways. On one side, insurers create a suite of plans including both capitation and FFS payment methods, which can then induce desired shares of patients under each by altering prices effectively through offerings. On the other side, instead of a market-based mechanism, practice managers adjust contracts at a regular time interval using historical knowledge of budget shortfalls and excesses. From this framework, doctors can recommend more or fewer visits to FFS patients for checkups, which patients tend to follow per Say's Law [7]. Hence, the timing of patient visits can be shifted by adjusting waiting times. Further, capitation patients can be preferentially recommended virtual visits (via email, phone, etc. with nurse practitioners as opposed to doctors). Our basic assumptions

This work was supported by the National Science Foundation [grant number DGE-1656518] and Stanford EDGE Fellowship. Any opinion, findings, and conclusions or recommendations expressed in this material are those of the author and do not necessarily reflect the views of the National Science Foundation.

include that practices do not turn patients away in order to maintain a certain capitation-to-FFS visit ratio, and that patients do not tend to switch between the payment systems at a meaningful level.

## II. METHODS

### A. Data

We use historical data to realistically estimate the coefficients in each of the two relevant models: insurer net cost and practice revenue. These US data are culled from several sources. First, data on counts of patients and patient visits, capitation and FFS revenues, and doctor and nurse salary and benefit costs are found in the 2014 Medical Group Management Association (MGMA) data [2]. Second, an estimate of annual capitation patient visits is derived from prior work showing 1.2 additional total visits per enrollee year for post-HMO (relative to pre-HMO) patients [8]. Third, hospitalization cost to insurers for FFS patients is the product of the number of days in an average hospital stay as of 2012 [9] and the per diem cost of an average FFS patient's hospital stay as of 2014 [1], i.e., 4.5 times $2,212. Fourth, the decrease in hospitalization cost when using a capitation payment system instead of FFS is derived from a 2011 study [10] finding a reduction of $7,679 per 1,000 member months.

All relevant variables (used to calculate $f_1$ and $f_2$ equilibria) are defined and listed with data estimates in Table 1. Henceforth, the term "capitation patient" refers to a patient under the capitation payment system who is expected to receive proactive team and nonvisit care during patient visits (whether in-person or virtual). The term "FFS patient" refers to a patient under the FFS payment system who is expected to receive traditional doctor care. All estimated values are for one year each.

### B. Modeling Insurer Net Cost

In our model, the goal of the insurer is to minimize their net cost, comprised of five components:

1. Annual FFS Cost: $(f_1 \times p) \times (f_2 \times n_f) \times r_f$ denotes the number of FFS patients, multiplied by the number of FFS patient visits per FFS patients, multiplied by the FFS net cost per FFS patient visit.
2. Annual Capitation Cost: $[(1-f_1) \times p] \times r_c$ denotes the number of capitation patients, multiplied by the capitation net cost per patient.
3. Annual Hospitalization Cost from FFS Patients: $(f_1 \times p) \times h_f$ denotes the number of FFS patients multiplied by the hospitalization cost per FFS patient.
4. Annual Hospitalization Cost from Capitation Patients: $[(1-f_1) \times p] \times h_c$ denotes the number of capitation patients multiplied by the hospitalization cost per capitation patient.
5. Performance-based Bonus (or Penalization): $\phi(z,\alpha,\xi)$ denotes the performance-based dollar amount paid by the insurer to the practice, where $z$ is the practice's performance metric (which is a function of $f_1$ and $f_2$), and $\alpha$ and $\xi$ are parameters to be set by the insurer.

TABLE I. VARIABLE DEFINITIONS AND ESTIMATED ANNUAL VALUES

| Variable | Description | Annual Estimated Value |
|---|---|---|
| $f_1$ | Fraction of FFS Patients | (Insurer-set Parameter) |
| $f_2$ | Fraction of FFS Patient Visits | (Practice-set Parameter) |
| $n_f$ | Number of Visits per FFS Patient [2] | 2.24 |
| $n_c$ | Number of Visits per Capitation Patient [2],[8] | 3.44 |
| $p$ | Number of Patients per Practice [2] | 1,684 |
| $r_f$ | FFS Revenue per Patient Visit [2] | $140.41 |
| $r_c$ | Capitation Revenue per Patient [2] | $346.32 |
| $h_f$ | Hospitalization Cost to Insurers per FFS Patient [1],[9] | $9,954.00 |
| $h_c$ | Hospitalization Cost to Insurers per Capitation Patient [1],[9]-[10] | $9,861.85 |
| $h_\varepsilon$ | $h_f - h_c$ | $92.15 |
| $c_d$ | Cost to Insurers for FFS Visit (Doctor) [2] | $63.56 |
| $c_n$ | Cost to Insurers for Capitation Visit (Nurse) [2] | $24.04 |
| $\alpha$ | Slope of Performance-Based Bonus | (Insurer-set Parameter) |
| $\xi$ | Cut-off Boundary of Performance-Based Bonus | (Insurer-set Parameter) $\to \infty$ |
| $z(f_1,f_2)$ | Practice Performance Metric | (Model-defined) |
| $\phi(z,\alpha,\xi)$ | Performance-Based Bonus (Paid by Insurer to Practice) | (Model-defined) |

The goal of setting a performance-based bonus function is to allow for insurers to incentivize practices to perform better services, so as to ameliorate the concern of capitation payment systems leading to lower quality of care. We first define the nonlinear performance-based bonus function $\phi(z,\alpha,\xi)$ model formulation, and then identify two constraints in defining $z$, the practice's performance metric. Let $\phi(z,\alpha,\xi)$ be a piecewise function equaling: $\alpha\xi$ if $z > \xi$, $\alpha z$ if $z \in [-\xi,\xi]$, and $-\alpha\xi$ if $z < -\xi$. As such, the performance-based bonus will be a dollar multiplier $\alpha$ of the practice's performance metric $z$ within bounds defined by parameter $\xi$. The existence of the $\xi$ parameter [6] allows the insurer to protect themselves against extraordinarily large performance-based payouts.

The first constraint on defining the practice's performance metric $z$ is to determine the spread of the function's output. Prior work has shown that capitation patients correspond to a "reduction in ambulatory-care sensitive ED visits of approximately 0.7 per 1,000 member months or approximately 22.6%" [10]. We assume symmetry, i.e. that FFS patients have a 22.6 percentage point increase in ambulatory-care sensitive ED visits relative to capitation patients. Hence, we define the practice performance metric $z(f_1, f_2)$ such that $z(0,0) = 0.113$ and $z(1,1) = -0.113$, so that the difference between a fully FFS system and a fully capitation system is 22.6 percentage points as found in the prior literature.

The second constraint on defining $z$ is that it does not scale linearly in $f_1$ or $f_2$ in order to obtain non-extreme equilibrium values (i.e., neither insurer nor practice will set the share of FFS patients or FFS patient visits to exactly 0 or 1). To this end, it is reasonable to envision a case where, with twice as many patient visits under FFS (presumably non-proactive) care, the quality of care diminishes more than twice as fast, due to physician burnout [11] and scaling factors for physicians working in teams [12]. Hence, we will assume a squared relationship in $f_2$. Further, it is generally assumed that patients under FFS tend to be healthier than those under capitation payments; so, it is also reasonable to assume that the quality of care diminishes less than twice as fast if we have double the share of patients treated under FFS. Specifically, capitation patients are often on Medicare, and hence likely older and sicker, whereas FFS patients are typically employed [13]. As such, we assume a square-root relationship in $f_1$. Combining, we define:

$$z(f_1, f_2) = -0.113 \sqrt{f_1} f_2^2 + 0.113(1 - \sqrt{f_1} f_2^2).$$

Finally, we sum and simplify the convex minimization problem for the insurer to solve:

$$\min_{f_1} f_1 p(f_2 n_f r_f - r_c + h_\epsilon) + \phi(z, \alpha, \xi) \; s.t. \; 0 \leq f_1, f_2 \leq 1$$

*C. Modeling Practice Revenue*

In our model, the goal of the practice is to maximize their profit, comprised of five components:
1. Annual FFS Revenue: $(f_1 \times p) \times (f_2 \times n_f) \times r_f$ as defined in the insurer net cost model.
2. Annual Capitation Revenue: $[(1-f_1) \times p] \times r_c$ as defined in the insurer net cost model.
3. Cost of Doctor from FFS Patients: $(f_1 \times p) \times (f_2 \times n_f) \times c_d$ denotes the number of FFS patients, multiplied by the number of FFS patient visits per FFS patients, multiplied by the mean practice cost (i.e. doctor income) per FFS patient visit.
4. Cost of Nurse from Capitation Patients: $[(1-f_1) \times p] \times [(1-f_2) \times n_c] \times c_n$ denotes the number of capitation patients multiplied by the mean practice cost (i.e. nurse income) per capitation patient visit.
5. Performance-based Bonus (or Penalization): $\phi(z,\alpha,\xi)$ as defined in the insurer net cost model.

We sum and simplify the convex maximization problem for the practice to solve:

$$\max_{f_2} f_2 p(n_f f_1 (r_f - c_d) + n_c c_n (1 - f_1)) + \phi(z, \alpha, \xi) \; s.t. \; 0 \leq f_1, f_2 \leq 1$$

*D. Playing the Stackelberg Game*

In an economic Stackelberg game, two players (a leader and a follower) take turns competing on quantity. In our analogous setting, the insurer and healthcare provider take turns effectively setting quantities (the number of patients under FFS, and the number of patient visits under FFS, respectively). In this paper, we assume only one insurer and one practice play the Stackelberg game; it is non-trivial to expand to multiple players in this framework. In one round of playing, the insurer plays first by setting $f_1$, and the practice responds by setting $f_2$, thus concluding the game. In a multiple-round game (also referred to as a "repeated game"), the insurer will solve the current round's minimization problem using the previous round's value of $f_2$ set by the practice, and likewise for the practice with the previous value of $f_1$. When multiple rounds are played, the insurer determines all actions (i.e., the values of $f_1$ to set) prior to the start of the Stackelberg game; similarly, the practice determines their actions over all rounds of the game in response to the insurer's actions.

The game is solved using basic backwards induction. Let the insurer's minimization expression derived above be defined as $min_{f_1} P(f_1, f_2)$ and the practice's maximization expression be $max_{f_2} \Pi(f_1, f_2)$. Since the insurer moves first by setting $f_1$, we define the best response for the practice as $R(f_1) = argmax_{f_2} \Pi(f_1, f_2)$. Given that the insurer can calculate how the practice will react, the best response for the insurer will be to play $f_1 = argmin_{f_1} P(f_1, R(f_1))$.

Using this method, we perform case analysis on the two types of solutions resulting from our choice of $\xi$: when $z \notin [-\xi,\xi]$, and when $z \in [-\xi,\xi]$. We summarize results for a one-round game, which are also applicable to multiple rounds if we can ensure that the value of $f_2$ given the previous $f_1$ value will be either $z \in [-\xi,\xi]$ for all rounds, or $z \notin [-\xi,\xi]$ for all rounds.

When $z \notin [-\xi,\xi]$, the practice will solve a linear optimization problem since $\phi(z,\alpha, \xi)$ is defined entirely by constants $\alpha$ and $\xi$ (and not $z$). Using the constant variables defined in Table 1, the insurer knows that the practice will choose $f_2$ to maximize $f_2 p(n_f f_1 (r_f - c_d) + n_c c_n (1 - f_1))$, yielding an equilibrium of $f_2 = 1$ since all other coefficient multipliers are known to be positive. In response, the insurer tries to minimize $f_1 p(f_2 n_f r_f - r_c + h_\epsilon)$. Now, substituting for the relevant expected values yields a positive coefficient on $f_1$, which necessitates that the insurer will set $f_1$ to 0. This pairing of equilibrium settings is non-optimal, as there is no incentive for health practices to take any patient visits under a capitation payment system, even when insurers have entirely capitation patients. As a counterexample, with $f_2 = 0.8$, the insurer's incentive flips so that $f_1 = 1$ and both players are better off. However, even this case is not conducive to incentivizing any patients to be under the capitation payment system. We note that if our value of $h_\varepsilon$ were less than \$31.80 (i.e., if there were a lower difference between hospitalization costs for FFS and capitation patients), then the equilibrium would be found at $f_1 = f_2 = 1$, and neither insurers nor practices would be incentivized to promote capitation payment systems.

In the case where $z \in [-\xi,\xi]$ (or if we simply assume that $\xi$ goes to infinity in our model), we solve both the insurer net cost minimization and the practice revenue maximization using $\phi(z, \alpha, \xi) = \alpha(0.113 - 0.226 \sqrt{f_1} f_2^2)$. This results in the practice setting $f_2 = [p/(0.452\alpha)] \cdot (\sqrt{f_1}(n_f(r_f - c_d) - n_c c_n) + n_c c_n / \sqrt{f_1})$. In this case, it is possible to find non-extreme (i.e., neither 0 nor 1) settings of $f_1$ and $f_2$ due to the nonlinearity of the performance-based bonus function $\phi(z,\alpha,\xi)$; these equilibria are more realistic to expect.

## III. RESULTS

To focus on the more interesting non-linear case where $z \in [-\xi,\xi]$, we let $\xi$ go to infinity. We can numerically find the equilibria for any given number of rounds of the Stackelberg game, since our alternating minimizing (insurer net cost) and maximizing (practice revenue) functions are both convex. Our model is able to find Stackelberg equilibria such that both insurers and practices will set non-extreme values for the share of FFS patients and FFS patient visits (hence indicating some incentive of switching to capitation payments). However, these equilibria yield a negative practice performance as defined by $z(f_1, f_2)$.

An example of a reasonable insurer choice is setting $\alpha = 682,000$ in a game played with any known number of alternating rounds. For only one round, we have $f_1 = 0.9536$ and $f_2 = 0.9397$; for 100 rounds, we have a very similar $f_1 = 0.9527$ and $f_2 = 0.9398$. Both settings of rounds result in a performance-based penalty for practices, which would incentivize the use of capitation payments; for one round, the insurer would earn an additional $-\phi = -\alpha \cdot z = -682,000 \cdot (0.113 - 0.226 \cdot \sqrt{0.9536} \cdot 0.9397^2) = \$55,843.00$ for that year from that practice. Similarly, for 100 rounds, the insurer would earn an additional \$55,808.54 annually.

Results are plotted in Fig 1 based on a game with one round and a game with 100 rounds (which appear representative of any number of rounds, both even and odd). The non-extreme ranges for $f_1$ and $f_2$ are intuitive since the insurer aims to minimize the performance-based bonus value (along with the original net cost), so it is reasonable that the $\alpha$ values chosen will be near the minimum of the performance-based bonus function plotted in red. Note that, based on this model, an equilibrium wherein both $f_1$ and $f_2$ are not set to extreme values necessarily results in a performance-based penalty as opposed to a bonus. One interpretation of this phenomenon is that in order to incentivize non-extreme settings of FFS versus capitation payments, practice performance would be sacrificed.

We comment that these results are robust to including revenue inflation in the model, which is currently used as an incentive for practices to convert to capitation payments over FFS. However, since capitation revenue and FFS revenue would likely be inflated year-over-year at a similar rate, we did not find strongly observable effects in causing higher shares of capitation patients or patient visits.

## IV. DISCUSSION

Our model formulation finds equilibria wherein both insurers and practices are incentivized to embrace the capitation system to some degree; however, these equilibria may not be the best option available to the involved patients. Our results are directly interpretable regarding the choice of capitation versus FFS patient and patient visit shares to be set by the involved parties in a Stackelberg game. This is the first model showing that, theoretically, both insurers and practices can have some degree of flexibility in setting usage of both FFS and capitation payment systems simultaneously. Further, finding the ensuing non-extreme equilibria between FFS and capitation payment systems implies that it is possible for both payment systems to co-exist within the current healthcare ecosystem, allowing for diversity of patient choice (though potentially at the expense of quality of patient care). These findings may be of use in future policy-based discussions on the benefits of shifting from FFS to capitation payments, and in particular show that it is possible for an insurer and practice to attempt a transitory phase having both payment systems in place.

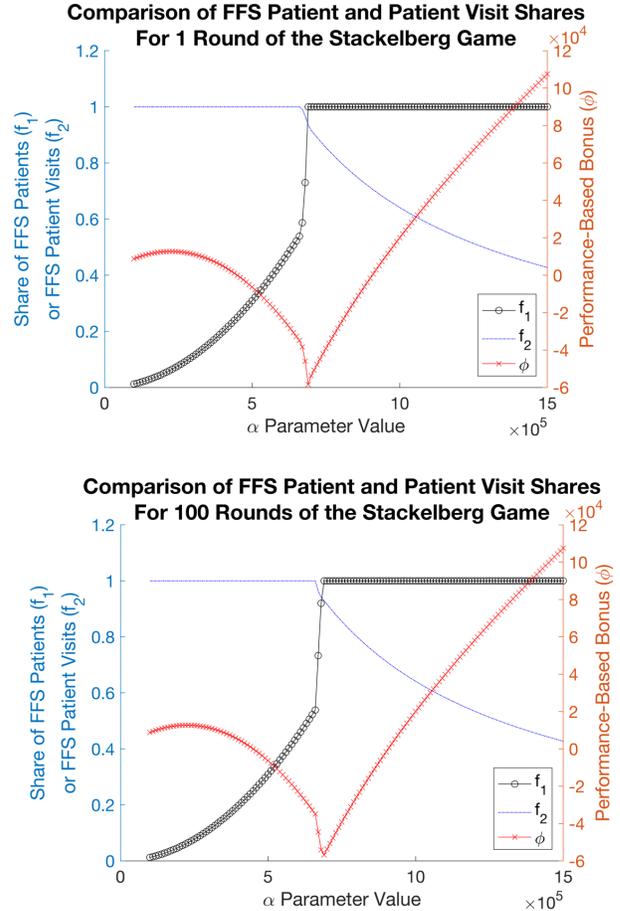

Fig. 1. (A) Top: Practice performance-based bonus/penalty parameter $\alpha$, set by the insurer, is shown against corresponding $f_1$ and $f_2$ values (black and blue lines, respectively) on the left axis, and resulting practice performance-based bonus on the right axis (red line), after playing one round of the Stackelberg game. There are no equilibria where neither $f_1$ nor $f_2$ have extreme values while also yielding a positive performance-based bonus. (B) Bottom: Similar results are shown after playing 100 rounds of the Stackelberg game.

The fact that non-extreme equilibria result from our Stackelberg game is due to our introduction of a non-linear performance-based bonus function; this takes into account that practice performance may decrease superlinearly when confronted with more FFS patient visits, and sublinearly when confronted with more (presumably healthier) FFS patients. With either a linear, or nonexistent, performance-based bonus function, both the alternating minimization and maximization functions for insurers and practices would be linear in their respective $f_1$, $f_2$ values, and would result in extreme solutions of either 0 or 1 for each setting of FFS patient shares and FFS patient visit shares. Instead, we have built a more realistic model wherein both payment systems can be implemented by insurers and practices.

We have hence shown that a reasonable payment mechanism can occur when the insurer sets the performance-based bonus using variable $\alpha$ to minimize their own net cost in the game, though this occurs at the expense of overall practice performance. Future work involves extending the model to multiple insurers and practices (similar to the Burdett-Shi-Wright model) and incorporating the direct relationships between each practice and individual insurers. Further, it could be useful to research patient behavior with regard to utility derived from FFS versus capitation payment systems; this line of work could allow for estimating the magnitude of second-order effects, such as insurer revenue, if incorporated in our Stackelberg game.

ACKNOWLEDGMENTS

We are grateful to Sanjay Basu and Mark Duggan for helpful discussions, and Ron Estrin and Fangdi Sun for proofreading. Any remaining errors are our own.